\documentclass[conference]{IEEEtran}
\usepackage{amssymb}
\usepackage{algorithm}
\usepackage{algpseudocode}
\usepackage{algorithmicx}
\usepackage{amsfonts}
\usepackage{amsmath}
\usepackage{amssymb}
\usepackage{amsthm}
\usepackage{array}
\usepackage{bbding}
\usepackage{bigints}
\usepackage{booktabs}
\usepackage{cite}
\usepackage{cleveref}
\usepackage{color}
\usepackage{diagbox}
\usepackage{epsfig,latexsym}
\usepackage{epstopdf}
\usepackage{graphicx}
\usepackage{fancyhdr}
\usepackage{float}
\usepackage{indentfirst}
\usepackage{lastpage}
\usepackage{makecell}
\usepackage{mathtools}
\usepackage{multirow}
\usepackage{pifont}
\usepackage{psfrag}
\usepackage{setspace}
\usepackage{stfloats}
\usepackage{empheq}
\usepackage{enumerate}
\usepackage{times}
\usepackage{subfigure}
\newcolumntype{C}{>{\centering\arraybackslash$}p{\linewidth}<{$}}
\newcounter{eqncnt}

\newcounter{eqnback}

\newtheorem{theorem}{Theorem}

\newtheorem{lemma}{Lemma}

\newtheorem{corollary}{Corollary}

\DeclareMathOperator{\diag}{diag}
\DeclareMathOperator{\tr}{tr} 

\begin{document}
\title{\Large STAR-RIS-Aided Cell-Free Massive MIMO with Imperfect Hardware\\
\vspace{-1em}
}

\author{\IEEEauthorblockN{Zeping Sui, Hien Quoc Ngo, and Michail Matthaiou}
\IEEEauthorblockA{Centre for Wireless Innovation (CWI), Queen's University Belfast, U.K. \\E-mail: zepingsui@outlook.com, \{hien.ngo, m.matthaiou\}@qub.ac.uk}
\vspace{-3em}}
\maketitle
\begin{abstract}
This paper considers a simultaneously transmitting and reflecting reconfigurable intelligent surface (STAR-RIS)-aided cell-free massive multiple-input multiple-output (CF-mMIMO) system, accounting for imperfect hardware in spatially correlated fading channels. Specifically, we consider the hardware impairments and phase noise at transceivers, as well as the phase shift errors generated within the STAR-RIS. We commence by introducing the STAR-RIS signal model, channel model, and imperfect hardware components. Then, the linear minimum mean-square error (MMSE) channel estimate is derived with pilot contamination, which provides sufficient information for sequential data processing. Moreover, a channel capacity lower bound is derived in the case of a finite number of RIS elements and access points (APs), while a closed-form expression for the downlink ergodic spectral efficiency (SE) for maximum ratio (MR) precoding is also deduced, where only the channel statistics are used. Our numerical results demonstrate that the STAR-RIS-aided CF-mMIMO system achieves higher SE compared to the conventional CF-mMIMO system, even with imperfect hardware. 
\end{abstract}

\section{Introduction}\label{Section 1}
Over the past years, cell-free massive multiple-input multiple-output (CF-mMIMO) has been widely investigated since it can mitigate the inter-cell interference of conventional cellular networks \cite{7827017,10640072}. Consequently, a CF-mMIMO system is capable of achieving better connectivity, enhanced spectral efficiency (SE), and improved energy efficiency (EE) compared to conventional cellular mMIMO networks \cite{zhang2020prospective}. However, the SE of CF-mMIMO systems may significantly deteriorate under harsh propagation scenarios with high path loss and correlated channels.

As a parallel development, reflecting reconfigurable intelligent surfaces (RISs) have emerged as an attractive technology to shape the transmit waveforms at the electromagnetic level without extra power amplifiers and digital signal processing methods, yielding passive beamforming and effective energy saving \cite{huang2019reconfigurable}. Therefore, RISs have been combined with CF-mMIMO systems, resulting in improved SE in harsh propagation channels \cite{van2021reconfigurable}. Particularly, both the uplink and downlink SEs of RIS-aided CF-mMIMO systems with single-antenna access points (APs) were investigated based on maximum ratio (MR) processing in \cite{van2021reconfigurable}. Then, the authors of \cite{10225319} carried out a performance analysis of RIS-aided CF-mMIMO systems with realistic transceiver hardware impairments (THIs), where uncorrelated channels and ideal RISs were considered. In \cite{10167480}, the uplink SE performance of RIS-aided CF-mMIMO systems with electromagnetic interference (EMI) was conceived. Nonetheless, the APs and user equipments (UEs) are located in the same reflection space in the above-mentioned RIS-aided CF-mMIMO systems, yielding a compromised system capacity. To this end, the concept of simultaneously transmitting and reflecting reconfigurable intelligent surface (STAR-RIS) was recently proposed to support full space coverage upon controlling the phases and amplitudes of transmitted waves \cite{9570143,10550177}. Consequently, the SE analysis of STAR-RIS-aided CF-mMIMO with ideal hardware was studied in \cite{10297571}. Then, beamforming schemes were designed to attain the maximum weighted sum rate in STAR-RIS-aided CF networks in \cite{10316600}. However, realistic imperfect hardware may lead to severe performance loss, which was ignored in the above-mentioned STAR-RIS-aided-CF-mMIMO-related papers. To this end, we hereafter investigate the performance of STAR-RIS-aided CF-mMIMO systems with realistic imperfect hardware under spatially correlated channels. The contributions of our paper are listed as follows:
\begin{itemize}
	\item We first derive the linear MMSE estimate of the cascaded channel by considering THIs, RIS phase errors and phase noise, where pilot contamination is considered.
	\item We derive a closed-form expression for the downlink ergodic SE of the STAR-RIS-aided CF-mMIMO system with MR precoding, imperfect hardware, and spatially correlated channels. Numerical results demonstrate that the proposed STAR-RIS-aided CF-mMIMO system can achieve significant SE improvement compared to conventional CF-mMIMO.
	\item The effect of different system parameters is analyzed, including transceiver hardware quality factors, the number of APs and the concentration parameter associated with RIS phase errors. It can be observed that higher hardware quality factors yield higher ergodic SE and lower channel estimation error, while the STAR-RIS-aided CF-mMIMO system still outperforms the conventional counterpart even in the worst RIS phase error case. Moreover, the value of the receiver hardware quality factor, $\gamma_R$, can affect the SE more significantly compared to the transmitter hardware factor $\gamma_T$. Finally, the accuracy of our derived closed-form expression of SE is also validated via Monte Carlo simulations. 
	\end{itemize}
\emph{Notation:} The expectation and trace operators are $\mathbb{E}\{\cdot\}$ and $\text{tr}(\cdot)$, respectively. We use $\mathcal{CN}(\pmb{a},\pmb{B})$ to denote a vector following a complex Gaussian distribution with mean $\pmb{a}$ and covariance matrix $\pmb{B}$; $(\cdot)^T$, $(\cdot)^H$ and $\left|\cdot\right|^2$ are the transpose, conjugate transpose and magnitude operators, respectively. $\pmb{I}_N$ denotes the $N$-dimensional identity matrix, while $I_m(\vartheta_p)$ represents the $m$-order Bessel function of the first kind with respect to $\vartheta_p$. Also, $\odot$, $\otimes$ and $\arg(\cdot)$ are the Hadamard product, Kronecker product and argument operators, respectively. The Euclidean norm operator is denoted by $\left \| \cdot \right \|$; $a^{(l)}$ and $A^{(m,n)}$ are the $l$th element of vector $\pmb{a}$ and the $(m,n)$th element of matrix $\pmb{A}$, respectively, while $\diag\{\pmb{a}\}$ and $\diag\{\pmb{A}\}$ denote a diagonal matrix constructed from the vector $\pmb{a}$ and from the main diagonal vector of matrix $\pmb{A}$, respectively. Finally, $\text{vec}(\pmb{A})$ denotes the vector obtained from stacking the columns of $\pmb{A}$. 
\section{System Model}\label{Section 2}
\subsection{STAR-RIS Signal Model and Operation Protocol}\label{Section 2-1}
\begin{figure}[t]
\centering
\includegraphics[width=0.8\linewidth]{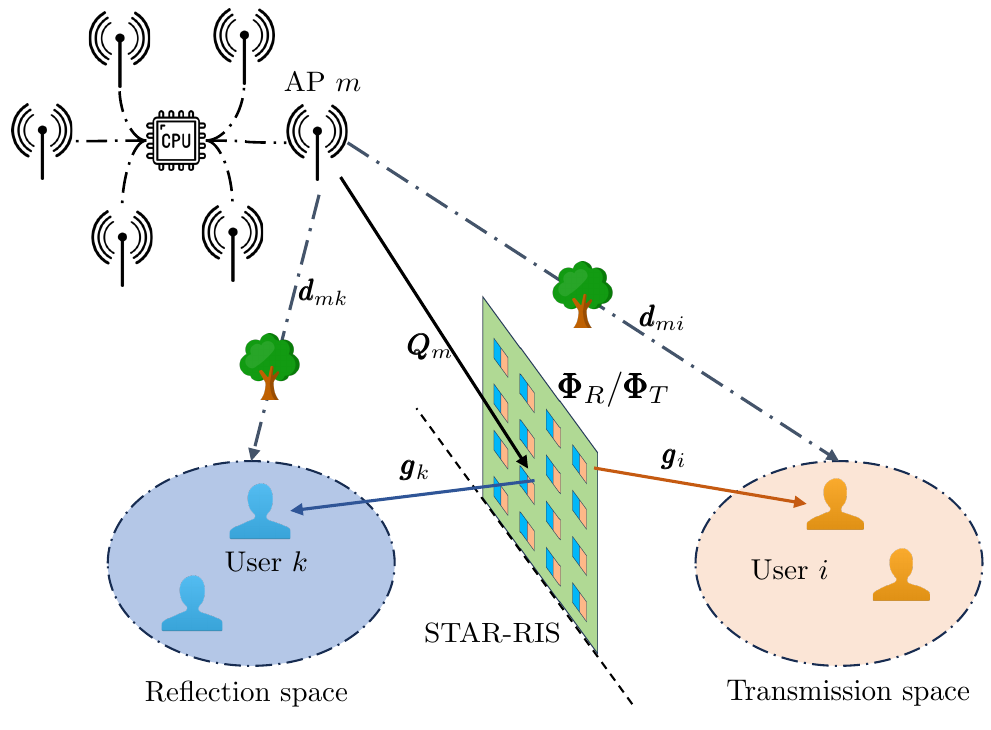}
\caption{Illustration of the considered STAR-RIS-aided CF-mMIMO system.}
\label{Figure1}
\vspace{-2em}
\end{figure}

As depicted in Fig. \ref{Figure1}, the considered STAR-RIS-aided CF-mMIMO system includes one STAR-RIS with $N$ elements, $M$ APs with $L$ antennas and $K$ UEs and utilizes the time-division duplexing (TDD) protocol. The overall space is partitioned into transmission and reflection spaces by the STAR-RIS. We denote $\mathcal{M}=\{1,2,\ldots,M\}$ as the set of APs, while the sets of STAR-RIS elements and UEs are denoted by $\mathcal{N}=\{1,2,\ldots,N\}$ and $\mathcal{K}=\{1,2,\ldots,K\}$, respectively. Based on the locations of UEs and the STAR-RIS, the UEs are divided into a reflection UE group $\mathcal{K}_T=\{1,2,\ldots,K_T\}$ and a transmission UE group $\mathcal{K}_R=\{1,2,\ldots,K_R\}$, i.e., we have $K=K_T+K_R$. Moreover, we assume that all APs that are connected to the CPU via error-free fronthaul links are randomly distributed in the reflection space, while the direct AP-UE links are blocked by obstacles. Given the $m$th AP's transmitted signal $\pmb{x}_m\in\mathbb{C}^{L}$, the reflected and transmitted signals of the $n$th STAR-RIS element can be formulated as $\sqrt{\beta_{T,n}}e^{j{\theta}_{T,n}}\pmb{x}_m$ and $\sqrt{{\theta}_{R,n}}e^{j{\theta}_{R,n}}\pmb{x}_m$, respectively, where ${\beta}_{T,n}$, ${\beta}_{R,n}$ and ${\theta}_{T,n}, {\theta}_{R,n}\in[0,2\pi)$ represent the amplitude and phase shift of the $n$th STAR-RIS element associated with the transmission and reflection spaces, respectively. Generally, phase shifts ${\theta}_{T,n}$ and ${\theta}_{R,n}$ are independent, while the amplitudes $\beta_{T,n}$ and $\beta_{R,n}$ are coupled based on the law of energy conservation as $\beta_{T,n}+\beta_{R,n}=1, \forall n\in\mathcal{N}$ \cite{9570143}. The STAR-RIS transmission- and reflection-coefficient matrices can be denoted as $\pmb{\Phi}_T=\text{diag}\left\{\sqrt{\beta_{T,1}}e^{j\theta_{T,1}},\ldots,\sqrt{\beta_{T,N}}e^{j\theta_{T,N}}\right\}$ and $\pmb{\Phi}_R=\text{diag}\left\{\sqrt{\beta_{R,1}}e^{j\theta_{R,1}},\ldots,\sqrt{\beta_{R,N}}e^{j\theta_{R,N}}\right\}$, respectively. In this paper, we consider the energy splitting (ES) operation protocol of STAR-RIS. All the STAR-RIS elements are set in reflection and transmission modes simultaneously. The STAR-RIS incident signal energy is divided into transmitted and reflected components based on $\beta_{T,n}$ and $\beta_{R,n}$. Hence, we have $\beta_{T,n},\beta_{R,n}\in[0,1]$, $\theta_{T,n}, \theta_{R,n}\in[0,2\pi)$.
\subsection{Phase Shift Errors}\label{Section 2-2}
We consider the realistic scenario that the phase shifts of the STAR-RIS elements are configured with finite precision, resulting in unavoidable phase errors \cite{papazafeiropoulos2021intelligent}. The phase errors can be modeled as $\tilde{\pmb{\Phi}}_k=\diag\left(e^{j\tilde{\theta}_{k,1}},\ldots,e^{j\tilde{\theta}_{k,N}}\right)$, where $\tilde{\theta}_{k,n}$ represents the phase error with $k\in\{T,R\}$ for $n=1,\ldots,N$. Moreover, the probability density function (PDF) of $\tilde{\theta}_{k,n}$ is symmetric with the zero-mean direction, hence, we have $\arg\left(\mathbb{E}\left[e^{j\tilde{\theta}_{k,n}}\right]\right)=0$, where $\arg(\cdot)$ denotes the argument operator. The phase noise elements are identically and independently distributed (i.i.d.) random variables (RVs) $\tilde{\theta}_{k,n}\in[-\pi,\pi)$ and follow the Von Mises distribution with zero mean and concentration parameter $\vartheta_p$, which indicates the phase estimation accuracy \cite{papazafeiropoulos2021intelligent}. The characteristic function of phase errors can be expressed as $\varsigma_p=I_1\left(\vartheta_p\right)/I_0\left(\vartheta_p\right)$, it should be noted that lower $\varsigma_p$ indicates a higher level of phase error \cite{papazafeiropoulos2021intelligent}.
\subsection{Channel Model}\label{Section 2-3}
Let us consider a quasi-static block fading channel with the coherence interval length, i.e., the number of channel uses of $\tau_c$. Moreover, $\tau_p$ channel uses are reserved for uplink channel estimation, while the length of downlink data transmission is $\tau_d=\tau_c-\tau_p$ channel uses. We consider that the APs and STAR-RIS as $L$-element uniform linear arrays (ULAs) and $(\sqrt{N}\times \sqrt{N})$-uniform squared planar array (USPA), respectively. As shown in Fig. \ref{Figure1}, the direct AP-UE link is given by $\pmb{d}_{mk}\sim\mathcal{CN}(\pmb{0},\pmb{R}^d_{mk})$, where $\pmb{R}^d_{mk}\in\mathbb{C}^{L\times L}$ denotes the spatial correlation matrix with the large-scale fading coefficient $\beta^d_{mk}=\tr(\pmb{R}^d_{mk})/L$. Moreover, we assume that the STAR-RIS is located close to the UEs, yielding RIS-UE line-of-sight (LoS) channel components. Consequently, the AP-RIS channel $\pmb{Q}_m\in\mathbb{C}^{L\times N}$ and RIS-UE channel $\pmb{g}_k\in\mathbb{C}^{N\times 1}$ can be respectively expressed as \cite{10297571}
\begin{align}\label{eq_nlos}
	{\pmb{Q}}_m=\sqrt{\xi_m}\pmb{R}_{\text{A},m}^{1/2}\pmb{V}_m\pmb{R}_\text{S}^{1/2},\ \pmb{g}_k=\sqrt{\frac{\alpha_k}{\iota_k+1}}\left(\sqrt{\iota_k}\bar{\pmb{g}}_k+\tilde{\pmb{g}}_k\right),
\end{align}
where $\xi_m$ and $\alpha_k$ denote the large-scale fading coefficients, $\iota_k$ is the Rician coefficient, $\pmb{R}_{\text{A},m}\in\mathbb{C}^{L\times L}$ and $\pmb{R}_\text{S}\in\mathbb{C}^{N\times N}$ are the deterministic Hermitian positive semi-definite spatially correlation matrices associated with the $m$th AP and the STAR-RIS, respectively, with $\text{vec}\left(\pmb{V}_m\right)\sim\mathcal{CN}(\pmb{0},\pmb{I}_{NL})$. Note that $\pmb{R}_{\text{A},m}$ can be obtained based on \cite{hoydis2013massive}. In addition, we consider an isotropic scattering with uniformly distributed multi-path transmission scenario, hence the elements of $\pmb{R}_\text{S}$ can be determined as $R_\text{S}(a,b)=d_H d_V\text{sinc}(2||\pmb{u}_{a}-\pmb{u}_{b}||^2/\lambda)$, and we have $\pmb{u}_l=[0,\mod(l-1,N_H)d_H,\lfloor(l-1)/N_H\rfloor d_V]^T$ with $l\in\{a,b\}$, where $d_H$ and $d_V$ are the horizontal width and the vertical height of each STAR-RIS element, while $N_H=\sqrt{N}$ is the number of RIS elements in each row \cite{bjornson2020rayleigh}. The LoS component $\bar{\pmb{g}}_k$ can be formulated as $\bar{\pmb{g}}_k=\pmb{a}_N(\varpi_k^a,\varpi_k^e)$, where $\varpi_k^a$ and $\varpi_k^e$ denote the the azimuth and elevation AoA from the STAR-RIS to UE $k$, respectively. The elements of the steering vector $\pmb{a}_{N}(\varpi_k^a,\varpi_k^e)\in\mathbb{C}^{N}$ can be given as
\begin{align}\label{eq_steering vector}
	{a}_{N}(\varpi_k^a,\varpi_k^e)(n)&=\exp\left\{j2\pi \frac{d_{\text{S}}}{\lambda}\left[\left \lfloor \frac{n-1}{\sqrt{N}}\right \rfloor\sin\varpi_k^a\sin\varpi_k^e\right.\right.\nonumber\\
	 &+\left.\left.\left((n-1)\mathrm{mod}\sqrt{N}\right)\cos\varpi_k^e\right]\right\},
\end{align}
where $\lambda$ denotes the wavelength, while $d_{\text{S}}$ is the spacing between the STAR-RIS elements for $n=1,\ldots,N$. Furthermore, the non-LoS (NLoS) component can be given as $\tilde{\pmb{g}}_k=\pmb{R}_{\text{S}}^{1/2}\pmb{c}_k\in\mathbb{C}^{N}$ with $\pmb{c}_k\sim\mathcal{CN}(\pmb{0},\pmb{I}_N)$. Consequently, the $L\times 1$-dimensional cascaded channel from the $m$th AP and the $k$th UE can be formulated as $\pmb{f}_{mk}=\pmb{d}_{mk}+\pmb{Q}_m\pmb{\Phi}_{k}\tilde{\pmb{\Phi}}_{k}\pmb{g}_k$, where $k\in\{T,R\}$ of $\pmb{\Phi}_{k}\tilde{\pmb{\Phi}}_{k}$ based on the location of UE $k$. Let $\pmb{G}_k=\bar{\pmb{g}}_k\bar{\pmb{g}}_k^H$, and the covariance matrix of the cascaded channel $\pmb{R}_{mk}=\mathbb{E}\{\pmb{f}_{mk}\pmb{f}_{mk}^H\}$ can be formulated as
\begin{align}\label{eq_R}	{\pmb{R}}_{mk}&=\pmb{R}^d_{mk}+\pmb{R}_{\text{A},m}\tr\left(\bar{\pmb{R}}^f_{mk}\right)+\pmb{R}_{\text{A},m}\tr\left(\tilde{\pmb{R}}^f_{mk}\right)\nonumber\\
	&=\pmb{R}^d_{mk}+\pmb{R}_{\text{A},m}\tr\left({\pmb{R}}^f_{mk}\right),
\end{align}
where we have
\begin{align}\label{eq_R_components}
	\bar{\pmb{R}}^f_{mk}=\frac{\xi_m\alpha_k\iota_k}{\iota_k+1}\pmb{R}_\text{S}{\pmb{\Phi}}_k\tilde{\pmb{G}}_k{\pmb{\Phi}}_k^H,\  \tilde{\pmb{R}}^f_{mk}=\frac{\xi_m\alpha_k}{\iota_k+1}\pmb{R}_\text{S}{\pmb{\Phi}}_k\tilde{\pmb{R}}_\text{S}{\pmb{\Phi}}_k^H,
\end{align}
where we have $\tilde{\pmb{G}}_k=\mathbb{E}\left\{\tilde{\pmb{\Phi}}_k\pmb{G}_k\tilde{\pmb{\Phi}}_k^H\right\}$ and $\tilde{\pmb{R}}_\text{S}=\mathbb{E}\left\{\tilde{\pmb{\Phi}}_k{\pmb{R}}_\text{S}\tilde{\pmb{\Phi}}_k^H\right\}$ with $\tilde{G}_{k}^{(n,n')}=G_{k}^{(n,n')}\mathbb{E}\left\{e^{j\tilde{\theta}_{k,n}-j\tilde{\theta}_{k,n'}}\right\}$ and $\tilde{R}_{\text{S}}^{(n,n')}=R_{\text{S}}^{(n,n')}\mathbb{E}\left\{e^{j\tilde{\theta}_{k,n}-j\tilde{\theta}_{k,n'}}\right\}$. Therefore, we can obtain $\tilde{\pmb{G}}_k=\varsigma^2_p{\pmb{G}}_k+(1-\varsigma^2_p)\diag({\pmb{G}}_k)$ and $\tilde{\pmb{R}}_\text{S}=\varsigma^2_p{\pmb{R}}_\text{S}+(1-\varsigma^2_p)\diag({\pmb{R}}_\text{S})$, where $\varsigma_p$ is the characteristic function of phase errors.
\vspace{-0.5em}
\subsection{Phase Noise Model}\label{Section 2-4}
We consider that both APs and UEs are equipped with free-running oscillators. The phase noise $\phi_m^\text{AP}(t)$ and $\phi_k^\text{UE}(t)$ terms associated with the channel use $t\in\{0,\ldots,\tau_c-1\}$ are generated during the baseband signal up-conversion and the passband signal down-conversion process. Explicitly, the phase noise samples can be modelled as a Wiener process, which can be formulated as \cite{bjornson2015massive}
\begin{align}\label{eq_wiener}
	\phi_m(t)&=\phi_m(t-1)+\zeta_{\phi_m}(t),\quad \zeta_{\phi_m}(t)\sim\mathcal{N}(0,\varrho^2_{\phi_m}),\nonumber\\
	\psi_k(t)&=\phi_k(t-1)+\zeta_{\psi_k}(t),\quad \:\:\zeta_{\psi_k}(t)\sim\mathcal{N}(0,\varrho^2_{\psi_k}),
\end{align}
where $\varrho^2_{\phi_m}$ and $\varrho^2_{\psi_k}$ are the variances associated with the additive terms $\zeta_{\phi_m}(t)$ and $\zeta_{\psi_k}(t)$. For the sake of analysis, we assume that $\phi_m(t)$ and $\psi_k(t)$ are i.i.d RVs across different UEs and APs with $\varrho^2_{\phi_m}=\varrho^2_{\phi}$ and $\varrho^2_{\psi_k}=\varrho^2_{\psi}$, $\forall k,m$. Specifically, the variances can be expressed as $\varrho^2_{i}=4\pi^2 f_c^2 c_iT_s$, where $f_c$ denotes the carrier frequency, $c_i$ is a constant associated with the oscillator with $i\in\{\phi,\psi\}$, while $T_s$ is the symbol duration \cite{bjornson2015massive}. Let $\varphi_{mk}(t)=\phi_m(t)+\psi_k(t)$ denote the overall oscillator phase noise. Hence, we have $\varphi_{mk}(t)\sim\mathcal{N}(\varphi_{mk}(t-1),\delta^2)$ with $\delta^2=\varrho^2_{\phi}+\varrho^2_{\psi}$. Consequently, the effective channel can be formulated as $\pmb{h}_{mk}(t)=\pmb{f}_{mk}e^{j\varphi_{mk}(t)}$.
\section{Channel Estimation}\label{Section 3}
We use $\tau_p<K$ orthogonal  pilots $\{\bar{\pmb{\phi}}_u\in\mathbb{C}^{\tau_p},u=1,\ldots,\tau_p\}$ in each coherence interval with $\|\bar{\pmb{\phi}}_u\|^2=\tau_p$, which are utilized by $K$ UEs. We define $\mathcal{P}_k\subset\{1,\ldots,K\}$ as the UE set that utilizes the same pilot symbol as the $k$th UE, including UE $u$ itself. Let $\bar{\pmb{\phi}}_{u_k}$ represent the pilot symbol used by UE $k$ with $u_k\in\{1,\ldots,\tau_p\}$. Hence, the pilot reuse pattern can be expressed as $\bar{\pmb{\phi}}^H_{u_k}\bar{\pmb{\phi}}_{u_i}=\tau_p,\ i\in\mathcal{P}_k$ and $\bar{\pmb{\phi}}^H_{u_k}\bar{\pmb{\phi}}_{u_i}=0,\ i\notin\mathcal{P}_k$. The time slots $t\in[0,\tau_p-1]$ are utilized for channel estimation. We assume the phase noise remains constant during channel estimation since we have $\tau_p\ll\tau_c$ \cite{7519076}. Due to the quantization noise introduced by I/Q imbalance and the Analog-to-Digital Converters (ADCs) of the receiver, we consider both transmitter and receiver distortion terms resembling realistic transceiver scenarios \cite{bjornson2017massive}. Particularly, we invoke the error vector magnitude (EVM) model to capture the THI \cite{10225319}. The received signal at the $m$th AP when $t=0$ can be formulated as
\begin{align}\label{eq_Z}
	\pmb{Z}_m(0)=\sqrt{\gamma_T}\sum_{i=1}^{K}\pmb{h}_{mi}(0)\pmb{x}^p_i+\pmb{G}_{m}^{\text{AP}}+\pmb{N}_m,
\end{align}
where $\pmb{x}^p_i=\sqrt{{p}\gamma_R}\bar{\pmb{\phi}}_{u_i}^H+\left(\pmb{w}_{i}^{\text{UE}}\right)^H$ and $\gamma_R,\gamma_T\in[0,1]$ are the hardware quality factors at the UE and the AP sides, respectively. Explicitly, $\gamma_R=\gamma_T=0$ denotes the worst hardware quality case, where there is no useful information signals transmitted from the UEs to APs, while $\gamma_R=\gamma_T=1$ represents the ideal hardware case. Moreover, ${p}>0$ denotes the transmitted power during channel estimation, while the elements of the noise term $\pmb{N}_m\in\mathbb{C}^{L\times \tau_p}$ are given by i.i.d. Gaussian RVs with zero mean and variance $\sigma^2$. The transmitter distortion is denoted as $\pmb{w}_{i}^{\text{UE}}\sim\mathcal{CN}(\pmb{0},(1-\gamma_R){p}\pmb{I}_{\tau_p})$, while the columns of the receiver distortion $\pmb{G}_{m}^{\text{AP}}$ can be formulated as $\pmb{g}_{m}^{\text{AP}}|\{\pmb{f}_{mi}\}\sim\mathcal{CN}\left(\pmb{0},(1-\gamma_T){p}\sum_{i=1}^{K}\diag\{\pmb{f}_{mi}\pmb{f}_{mi}^H\}\right)$. To estimate $\pmb{h}_{mk}(0)=\pmb{f}_{mk}e^{j\varphi_{mk}(0)}$, we correlate the received signal and the normalized pilot signal as $\pmb{z}_{mk}(0)\triangleq\pmb{Z}_m(0)\bar{\pmb{\phi}}_{u_k}/\sqrt{\tau_p}$, yielding
\begin{align}\label{eq_z}
	\pmb{z}_{mk}(0)&=\sum_{i\in\mathcal{P}_k}\sqrt{\kappa_p}\pmb{h}_{mi}(0)+\sqrt{\frac{\gamma_T}{\tau_p}}\sum_{i=1}^K\pmb{h}_{mi}(0)\left(\pmb{w}_{i}^{\text{UE}}\right)^H\bar{\pmb{\phi}}_{u_k}\nonumber\\
	&+\frac{\pmb{G}_{m}^{\text{AP}}\bar{\pmb{\phi}}_{u_k}}{\sqrt{\tau_p}}+\pmb{n}_{mk},
\end{align}
where $\kappa_p=\gamma_T\gamma_R{p}\tau_p$, $\pmb{n}_{mk}\triangleq{\pmb{N}_m\bar{\pmb{\phi}}_{u_k}}/{\sqrt{\tau_p}}\sim\mathcal{CN}(\pmb{0},\sigma^2\pmb{I}_L)$ and ${\pmb{G}_{m}^{\text{AP}}\bar{\pmb{\phi}}_{u_k}}/{\sqrt{\tau_p}}|\{\pmb{f}_{mi}\}$ obeys the same distribution as $\pmb{g}_{m}^\text{AP}|\{\pmb{f}_{mi}\}$. By exploiting the MMSE channel estimation, we can derive the estimated cascaded channel $\hat{\pmb{h}}_{mk}(0)$ as
\begin{align}\label{eq_ch_estimated}
	\hat{\pmb{h}}_{mk}(0)=\sqrt{\kappa_p}\pmb{R}_{mk}\pmb{\Psi}_{mk}^{-1}\pmb{z}_{mk}(0),
\end{align}
where $\pmb{\Psi}_{mk}=\kappa_p\sum_{i\in\mathcal{P}_k}\pmb{R}_{mi}+(1-\gamma_R)\gamma_T{p}\sum_{i=1}^K\pmb{R}_{mi}+(1-\gamma_T){p}\sum_{i=1}^K\diag(\pmb{R}_{mi})+\sigma^2\pmb{I}_L$. It should be noted that $\pmb{\Psi}_{mk}$ and $\pmb{R}_{mk}$ are deterministic Hermitian matrices. Consequently, for the estimated channel $\hat{\pmb{h}}_{mk}(0)$ and the estimation error $\tilde{\pmb{h}}_{mk}(0)=\pmb{h}_{mk}(0)-\hat{\pmb{h}}_{mk}(0)$, we have $\hat{\pmb{h}}_{mk}(0)\sim\mathcal{CN}({\pmb{0}},\pmb{\Omega}_{mk})$ and $\tilde{\pmb{h}}_{mk}(0)\sim\mathcal{CN}(\pmb{0},\pmb{C}_{mk})$ with $\pmb{\Omega}_{mk}=\gamma_T\gamma_R{p}\tau_p\pmb{R}_{mk}\pmb{\Psi}_{mk}^{-1}\pmb{R}_{mk}$ and $\pmb{C}_{mk}=\pmb{R}_{mk}-\pmb{\Omega}_{mk}$.
\vspace{-1em}
\section{Downlink Data Transmission}\label{Section 4}
\vspace{-1em}
To derive the closed-form expression of the downlink SE, we utilize the MR precoding scheme in this paper rather than the minimum mean square error (MMSE) precoding scheme. Moreover, the MR precoding has lower complexity compared to the MMSE precoding. During the downlink data transmission time slot $t\in[\tau_p,\tau_c-1]$, the estimated channels $\hat{\pmb{h}}_{mk}(0)$ are utilized to formulate MR precoding vectors. Upon considering the transmitter hardware distortion $\pmb{\mu}_{m,t}$, the downlink transmitted signal of AP $m$ can be expressed as $\pmb{x}_m(t)=\bar{\pmb{x}}_m(t)+\pmb{\mu}_{m}^\text{AP}(t)$, where $\bar{\pmb{x}}_m(t)=\sqrt{\gamma_T\rho}\sum_{k=1}^K\sqrt{\eta_{mk}}\hat{\pmb{h}}_{mk}(0)s_k(t)\in\mathbb{C}^L$ where $\rho$ denotes the normalized downlink SNR, $s_k$ is the transmitted symbol of UE $k$ with $\mathbb{E}\{|s_k(t)|^2\}=1$, and $\eta_{mk}$ represents the corresponding power control coefficient. For the conventional power control scheme, we assume that the power control coefficient of AP $m$ keeps constant when associated with different UEs, i.e., $\eta_{mk}=\left(\sum_{i=1}^K\tr(\pmb{\Omega}_{mi})\right)^{-1},\forall m,k$. We also define the MR precoding vector set $\left\{\hat{\pmb{h}}_{mk}(0),\forall k\right\}$. The transmitter hardware distortion $\pmb{\mu}_{m,t}$ can be formulated as $\pmb{\mu}_{m}^\text{AP}(t)\sim\mathcal{CN}\left(\pmb{0},\pmb{D}_{m,\left\{\hat{\pmb{h}}\right\}}\right)$ \cite{bjornson2017massive}, where the $L\times L$-dimensional conditional correlation matrix can be expressed as $\pmb{D}_{m,\left\{\hat{\pmb{h}}\right\}}=(1-\gamma_T)\sum_{k=1}^K\rho\eta_{mk}\diag(\pmb{\Omega}_{mk})$, where $\diag(\pmb{\Omega}_{mk})$ denotes the main diagonal vector of $\pmb{\Omega}_{mk}$. It can be readily shown that we have the power constraint $\mathbb{E}\{||\bar{\pmb{x}}_m(t)||^2\}\leq\gamma_T\rho$, yielding $\sum_{k=1}^K\eta_{mk}\tr(\pmb{\Omega}_{mk})\leq 1$. The receiver distortion term is given by ${\mu}_{k}^\text{UE}(t)\sim\mathcal{CN}\left(0,(1-\gamma_R)\nu_{m,\{\hat{\pmb{h}}\}}\right)$ with
\begin{align}\label{eq_re_distortion}
	\nu_{m,\{\hat{\pmb{h}}\}}&=\mathbb{E}\left\{\left|\sum_{m=1}^M\pmb{h}_{mk}^H(t)\pmb{x}_m(t)\right|^2|\{\pmb{h}_{mk}(t)\}\right\}\nonumber\\
	&=\mathbb{E}\left\{\sum_{m=1}^M\sum_{i=1}^K\rho\eta_{mi}\left(\gamma_T|\pmb{h}_{mk}^H(t)\hat{\pmb{h}}_{mi}(0)|^2\right.\right.\nonumber\\
	 &+\left.\left.(1-\gamma_T)\left \|\hat{\pmb{h}}_{mi}(0)\odot\pmb{h}_{mk}(t)\right \|^2\right)\right\}.
\end{align}
Consequently, the received signal at UE $k$ can be formulated as ${y}_k(t)$ in \eqref{eq_received_y} shown at the top of next page, where $\kappa_\rho=\gamma_T\gamma_R\rho$, $\tilde{w}_k(t)={\mu}_{k}^\text{UE}(t)+{n}_k(t)$ and $n_k(t)\sim\mathcal{CN}(0,1)$ denotes the noise term. Moreover, we have
\setcounter{eqnback}{\value{equation}} \setcounter{equation}{9}
\begin{figure*}[!t]
\begin{align}\label{eq_received_y}
	{y}_k(t)&=\sum_{m=1}^M \sqrt{\gamma_R}\pmb{h}_{mk}^H(t)\pmb{x}_m(t)+{\mu}_{k}^\text{UE}(t)+{n}_k(t)\nonumber\\	
	&=\sum_{m=1}^M \sqrt{\gamma_R}\pmb{h}_{mk}^H(t)\left(\sqrt{\gamma_T\rho}\sum_{i=1}^K\sqrt{\eta_{mi}}\hat{\pmb{h}}_{mi}(0)s_i(t)+\pmb{\mu}_{m}^\text{AP}(t)\right)+{\mu}_{k}^\text{UE}(t)+{n}_k(t)\nonumber\\	
	&=\sqrt{\kappa_\rho}\sum_{m=1}^M\sqrt{\eta_{mk}}\pmb{h}_{mk}^H(t)\hat{\pmb{h}}_{mk}(0)s_k(t)+\sqrt{\kappa_\rho}\sum_{i\neq k}^K\sum_{m=1}^M\sqrt{\eta_{mi}}\pmb{h}_{mk}^H(t)\hat{\pmb{h}}_{mi}(0)s_{i}(t)+\sqrt{\gamma_R}\sum_{m=1}^M\pmb{h}_{mk}^H(t)\pmb{\mu}_{m}^\text{AP}(t)+\tilde{w}_k(t)\nonumber\\
	&={\tt DS}_k\cdot s_k(t)+{\tt BU}_k\cdot s_k(t)+\sum_{i\neq k}^K {\tt UI}_{ki}\cdot s_{i}(t)+{\tt HI}_{k}+{\mu}_{k}^\text{UE}(t)+{n}_k(t),	
\end{align}
\hrulefill
\vspace{-1.5em}
\end{figure*}
\setcounter{eqncnt}{\value{equation}}
\setcounter{equation}{\value{eqnback}}
\begin{align}\label{eq_DS}\setcounter{equation}{10}	{\tt DS}_k(t)\triangleq\sqrt{\gamma_R\gamma_T\rho}\mathbb{E}\left\{\sum_{m=1}^M\sqrt{\eta_{mk}}\pmb{h}^H_{mk}(t)\hat{\pmb{h}}_{mk}(0)\right\},
\end{align}
\vspace{-1em}
\begin{align}\label{eq_BU}
	{\tt BU}_k(t)&\triangleq\sqrt{\gamma_R\gamma_T\rho}\left(\sum_{m=1}^M\sqrt{\eta_{mk}}\pmb{h}^H_{mk}(t)\hat{\pmb{h}}_{mk}(0)\right.\nonumber\\
	&\left.-\mathbb{E}\left\{\sum_{m=1}^M\sqrt{\eta_{mk}}\pmb{h}^H_{mk}(t)\hat{\pmb{h}}_{mk}(0)\right\}\right),
\end{align}
\vspace{-1em}
\begin{align}\label{eq_UI}
	{\tt UI}_{ki}(t)\triangleq\sqrt{\gamma_R\gamma_T\rho}\sum_{m=1}^M\sqrt{\eta_{mi}}\pmb{h}^H_{mk}(t)\hat{\pmb{h}}_{mi}(0),
\end{align}
\vspace{-1em}
\begin{align}\label{eq_TH}
	{\tt HI}_{k}(t)\triangleq\sqrt{\gamma_R}\sum_{m=1}^M\pmb{h}^H_{mk}(t)\pmb{\mu}_{m}^\text{AP}(t),
\end{align}
denoting the strength of the desired signal (DS), beamforming uncertainty (BU), UE interference (UI) introduced by UE $i$ and the interference caused by transmitter hardware impairments (HI), respectively. Upon invoking the use-and-then forget capacity bound \cite{7827017}, the downlink ergodic SE (measured in bit/s/Hz) associated with UE $k$ can be formulated as
\begin{align}\label{eq_DL_SE}
	R_k=\frac{1}{\tau_c}\sum_{t=\tau_p}^{\tau_c-1}\log_2\left(1+{\tt SINR}_k(t)\right),
\end{align}
where the effective signal-to-interference-plus-noise ratio (SINR) of UE $k$ at the channel use $t$ can be expressed as
\begin{align}\label{eq_SINR}
	{\tt SINR}_k(t)=\frac{|{\tt DS}_k(t)|^2}{{\tt EN}_k(t)},
\end{align}
where ${\tt EN}_k(t)=\mathbb{E}\{|{\tt BU}_k(t)|^2\}+\sum_{i\neq k}^K\mathbb{E}\{|{\tt UI}_{ki}(t)|^2\}+\mathbb{E}\{|{\tt HI}_k(t)|^2\}+\mathbb{E}\{|{\mu}_{k}^\text{UE}(t)|^2\}+1$. Then, the sum SE can be expressed as $R = \sum_{k=1}^K R_k$.

\emph{Proposition 1:} Upon exploiting the MR precoding with arbitrary phase shift matrix $\pmb{\Phi}_k$, the closed-form expression for  downlink ergodic SE associated with the $k$th UE can be formulated as in \eqref{eq_DL_SE}, where
\begin{align}\label{eq_SE_closed}
	{\tt SINR}_k(t)=\frac{\gamma_R\gamma_T\rho e^{-\delta^2 t}\left|\tr(\pmb{\eta}_k^{1/2}\pmb{\Omega}_k)\right|^2}{J_k(t)},
\end{align}
where $J_k(t)$ is given by \eqref{eq_Ik}, shown at the top of the next page, while we introduce the following notations: $\tilde{\gamma}=(1-\gamma_R)(1-\gamma_T)$, $\pmb{\eta}_k^{1/2}=\pmb{P}_k^{1/2}\otimes\pmb{I}_L$ with $\pmb{P}_k^{1/2}=\diag\left(\sqrt{\eta_{1k}},\ldots,\sqrt{\eta_{Mk}}\right)$, $\pmb{\Omega}_k=\diag\left(\pmb{\Omega}_{1k},\ldots,\pmb{\Omega}_{Mk}\right)$, $\pmb{R}_k=\diag\left(\pmb{R}_{1k},\ldots,\pmb{R}_{Mk}\right)$, $\pmb{A}_k=\diag(a_{1k},\ldots,a_{Mk})$ with $a_{mk}=|\tr(\pmb{\Omega}_{mk})|^2$, $\pmb{B}_{ik}=\diag(b_{1ik},\ldots,b_{Mik})$ with $b_{mik}=\left|\gamma_T\gamma_R p\tau_p\tr(\pmb{R}_{mi}\pmb{\Psi}_{mk}^{-1}\pmb{R}_{mk})\right|^2$, and $\pmb{\Xi}_{ik}=\diag(\pmb{\Xi}_{1ik},\ldots,\pmb{\Xi}_{Mik})$ with $\pmb{\Xi}_{mik}=\pmb{R}_{mi}\pmb{R}_{mk}^{-1}$. The proof is omitted due to space constraints.
\setcounter{eqnback}{\value{equation}} \setcounter{equation}{17}
\begin{figure*}[!t]
\begin{align}\label{eq_Ik}
J_k(t)&=\gamma_T\gamma_R\rho e^{-\varrho^2_{\phi}t}\left(1-e^{-\varrho^2_{\psi}t}\right)\left|\tr(\pmb{\eta}_k^{1/2}\pmb{\Omega}_k)\right|^2+\gamma_T\rho\left(1-\gamma_Re^{-\varrho^2_{\phi}t}\right)\tr(\pmb{P}_k\pmb{A}_k)+\tilde{\gamma}\rho\tr\left(\pmb{\eta}_k(\diag(\pmb{\Omega}_{k}))^2\right)\nonumber\\
	&+\sum_{i\in\mathcal{P}_k\setminus\{k\}}\left\{\gamma_T\rho\left(1-\gamma_Re^{-\varrho^2_{\phi}t}\right)\tr(\pmb{P}_i\pmb{B}_{ik})+\gamma_T\gamma_R\rho e^{-\varrho^2_{\phi}t}\left|\tr\left(\pmb{\eta}_k^{1/2}\pmb{\Xi}_{ik}\pmb{\Omega}_k\right)\right|^2\right\}\nonumber\\
	&+\sum_{i=1}^K\rho\left[\gamma_T\tr(\pmb{\eta}_i\pmb{R}_k\pmb{\Omega}_i)+(1-\gamma_T)\tr\left(\pmb{\eta}_i\diag(\pmb{\Omega}_i)\diag(\pmb{R}_k)\right)\right]+1,
\end{align}
\hrulefill
\vspace{-1em}
\end{figure*}
\setcounter{eqncnt}{\value{equation}}
\setcounter{equation}{\value{eqnback}}
\section{Numerical Results}
In this section, we provide some simulation results to characterize the overall performance of our proposed STAR-RIS-aided CF-mMIMO system with imperfect hardware. Consider a geographic area with the size of $1\times 1$ $\text{km}^2$, where the locations of nodes are given in terms of $(x,y)$ coordinates in meter units. The STAR-RIS is located at the centre point with the coordinates of $(0,0)$, where the $x<0$ region denotes the reflection space and the transmission space is denoted by the $x>0$ region, $\forall y$. The $M$ APs are uniformly distributed in a squared region with $x_{\text{AP}}\in[-500,-250]$ and $y_{\text{AP}}\in[250,500]$. The $K_R$ UEs in the reflection space are uniformly distributed within a sub-region of $x_R,y_R\in[-325,-125]$, while the sub-region associated with the transmission space is denoted as $x_T\in[125,325]$ and $y_T\in[-125,-325]$. The carrier frequency is $f_c=2$ GHz, and the bandwidth is $10$ MHz. The length of the coherence block is $\tau_c=100$, while we utilize $\tau_p=3$ pilots to carry out channel estimation, which corresponds to a coherence bandwidth $W=100$ KHz and the length of the coherence block is $T_c=1$ ms. The power of the uplink transmitted pilot and the maximum downlink transmitted symbol power are given by $p=0.2$ W and $\rho=1$ W, respectively. The symbol duration can be expressed as $T_s=10$ $\mu s$, while the oscillator constant is set as $c_i=1\times 10^{-18},\forall i$. We deploy $K_R=K_T=3$ UEs in the reflection and transmission spaces, respectively, and $L=4$ antennas are deployed at every AP. Moreover, the horizontal width and the vertical height of each RIS element are $d_H=d_V=\lambda/4$. The RIS and AP element spacings are used as $d_{\text{RIS}}=\lambda/4$ and $d_{\text{AP}}=\lambda/2$, respectively. The heights of UEs, RIS and AP are $1.5$ m, $30$ m and $12.5$ m, respectively. The noise power is $\sigma^2=-97$ dBm, while the other channel-related parameters are the same as those in \cite{ozdogan2019performance}.
\begin{figure}[htbp]
\centering
\centering
\includegraphics[width=0.8\linewidth]{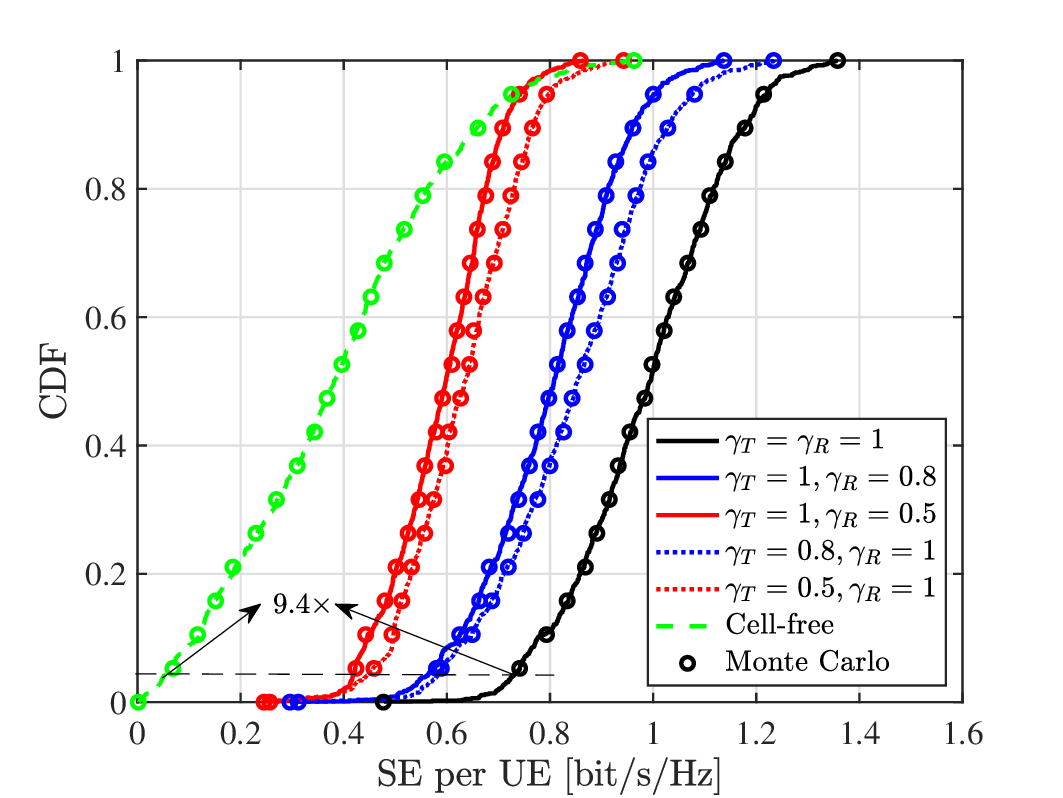}
\caption{CDF of the sum downlink SE for STAR-RIS-aided CF-mMIMO with different $\gamma_T$ and $\gamma_R$ and CF-mMIMO with ideal hardware ($M=20,K_R=K_T=3,N=128,L=4$).}
\vspace{-2em}
\label{Figure2}
\end{figure}

In Fig. \ref{Figure2}, we investigate the cumulative distribution function (CDF) of the downlink sum SEs corresponding to the conventional CF-mMIMO (denoted as ``Cell-free'') with ideal hardware and our proposed STAR-RIS-aided CF-mMIMO systems with imperfect hardware and $\vartheta_p=4$. The closed-form expression and Monte Carlo simulation results are calculated based on \eqref{eq_DL_SE}, \eqref{eq_SINR} and \eqref{eq_Ik}. Specifically, we consider different hardware quality factor settings of $\gamma_T$ and $\gamma_R$. Based on Fig. \ref{Figure2}, we have the following observations. Firstly, the STAR-RIS-aided CF-mMIMO systems can attain significant SE performance gains compared to the conventional CF-mMIMO system, regardless of the hardware quality of the former architecture. Secondly, with $\gamma_T=\gamma_R=1$, the $95\%$-likely SE of STAR-RIS-aided CF-mMIMO systems is about $14.7$ times higher than that of the CF-mMIMO systems. Moreover, given the value of $\gamma_T$ ($\gamma_R$), the smaller $\gamma_R$ ($\gamma_T$) yields the worse SE performance. This is because lower hardware quality unavoidably leads to a lower SE. Compared to the $(\gamma_T,\gamma_R)=(0.5,1)$ case, a lower SE can be attained when we use $(\gamma_T,\gamma_R)=(1,0.5)$. This observation can also be seen when we consider $(\gamma_T,\gamma_R)=(0.8,1)$ and $(\gamma_T,\gamma_R)=(1,0.8)$ settings. It implies that $\gamma_R$ has a more significant impact on the value of $I_k(t)$ in \eqref{eq_Ik}. Finally, we can observe a perfect overlap between the Monte Carlo simulations and the SE closed-form expression, which confirms the accuracy of our derived analytical results.

\begin{figure}[htbp]
\centering
\centering
\includegraphics[width=0.8\linewidth]{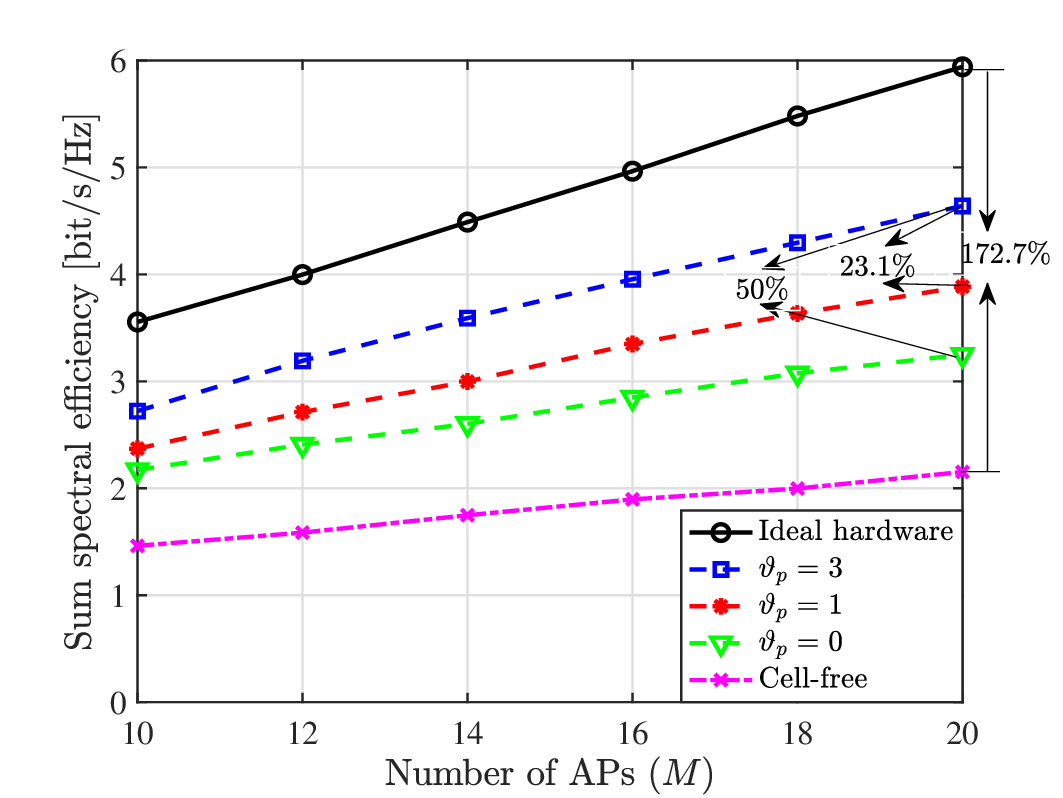}
\caption{Downlink sum SE versus the number of AP $M$ operating for different RIS phase error parameter $\vartheta_p$ ($K_R=K_T=3,N=64,L=4,\gamma_T=\gamma_R=1$).}
\label{Figure3}
\vspace{-2em}
\end{figure}

We now provide the sum SE of the STAR-RIS-aided CF-mMIMO with $\gamma_T=\gamma_R=1$ as a function of the number of APs, $M$, with different $\vartheta_p$ values, where the conventional CF-mMIMO system with ideal hardware is also depicted as a benchmark. From Fig. \ref{Figure3}, we can readily see that higher SEs can be attained by deploying more APs since more efficient beamforming can be achieved when we increase the number of spatial degrees of freedom. Moreover, the STAR-RIS-aided CF-mMIMO systems offer significant SE gains compared to the conventional CF-mMIMO system. Explicitly, the STAR-RIS-aided CF-mMIMO is capable of providing about $172.7\%$ higher sum SE compared to the CF-mMIMO under the phase error-free and $M=20$ scenario. Furthermore, the STAR-RIS-aided CF-mMIMO system with higher values of $\vartheta_p$ yields better sum SE performance since lower phase errors appear on the STAR-RIS. Specifically, given $M=20$, the sum SE of $\vartheta_p=3$ is capable of striking $23.1\%$ and $50\%$ gains compared to the $\vartheta_p=1$ and $\vartheta_p=0$ cases. It should be noted that the STAR-RIS-aided CF-mMIMO system always outperforms the CF-mMIMO counterpart even in the worst case of $\vartheta_p=0$.

\section{Conclusions}
In this paper, a STAR-RIS-aided CF-mMIMO system with imperfect hardware was conceived, where transceiver hardware impairments, phase noise and RIS phase errors were considered under spatially correlated channels. The linear MMSE estimate of the cascaded channel was derived to reduce the channel estimation overhead. Moreover, a closed-form expression for the ergodic downlink SE under MR precoding was derived, where both the pilot contamination and channel estimation errors were taken into account. The numerical results illustrated the substantial performance benefits of exploiting STAR-RIS in CF-mMIMO systems, compared to the conventional CF-mMIMO without STAR-RIS. The SE performance under different hardware parameter settings was also characterized.
\section*{Acknowledgement}
This work is a contribution by Project REASON, a UK Government funded project under the Future Open Networks Research Challenge (FONRC) sponsored by the Department of Science Innovation and Technology (DSIT).  The work of H. Q. Ngo was supported by the U.K. Research and Innovation Future Leaders Fellowships under Grant MR/X010635/1, and a research grant from the Department for the Economy Northern Ireland under the US-Ireland R\&D Partnership Programme. The work of M. Matthaiou was in part by the U.K. Engineering and Physical Sciences Research Council (EPSRC) under Grant EP/X04047X/1 and by the European Research Council (ERC) under the European Union's Horizon 2020 research and innovation programme (grant agreement No. 101001331).
\renewcommand{\refname}{References}
\mbox{} 
\nocite{*}
\bibliographystyle{IEEEtran}
\bibliography{ref.bib}
\end{document}